# Homotopic Gradients of Generative Density Priors for MR Image Reconstruction

Cong Quan, Jinjie Zhou, Yuanzheng Zhu, Yang Chen, Shanshan Wang, Dong Liang, *Senior Member, IEEE*, Qiegen Liu, *Senior Member, IEEE*

*Abstract*—Deep learning, particularly the generative model, has demonstrated tremendous potential to significantly speed up image reconstruction with reduced measurements recently. Rather than the existing generative models that often optimize the density priors, in this work, by taking advantage of the denoising score matching, homotopic gradients of generative density priors (HGGDP) are exploited for magnetic resonance imaging (MRI) reconstruction. More precisely, to tackle the low-dimensional manifold and low data density region issues in generative density prior, we estimate the target gradients in higher-dimensional space. We train a more powerful noise conditional score network by forming high-dimensional tensor as the network input at the training phase. More artificial noise is also injected in the embedding space. At the reconstruction stage, a homotopy method is employed to pursue the density prior, such as to boost the reconstruction performance. Experiment results implied the remarkable performance of HGGDP in terms of high reconstruction accuracy; only 10% of the k-space data can still generate image of high quality as effectively as standard MRI reconstructions with the fully sampled data.

## I. INTRODUCTION

Magnetic Resonance Imaging (MRI) noninvasively depicts structural and functional features inside a patient in rich contrasts, supporting today's medical diagnosis and researches. However, one major drawback of MRI is the relatively slow data acquisition speed, causing non-idealized space resolution, patient discomfort and hinders applications in time-critical diagnoses. Under the premise of guaranteeing the image quality, various acceleration techniques have been developed to speed up acquisition time.

Recently, the development of compressed sensing (CS) theory has shown that it is possible to reconstruct MR images from much fewer k-space (i.e. Fourier space) measurements [1]. Specifically, CS allows fast acquisition that bypasses the Nyquist-Shannon sampling criteria with more aggressive undersampling [2]. To apply the CS theory to MRI reconstruction, it must find a proper sparsifying transformation to make the signal sparse and solve a minimization with regularizers. Particularly, CS-MRI focuses on applying predefined sparsify transforms, such as wavelet transformation [3], the discrete cosine transform (DCT) [4], total variation (TV) [5], [6], or contourlet transform [7], [8], and developing efficient numerical algorithms to solve nonlinear optimization problems [9], [10]. Besides, CS-MRI has been extended to a more flexible sparse representation learned directly from data using dictionary learning, which expresses local features of reconstructed images more accurately compared to predefined universal transforms [11]-[14]. Even though dictionary learning-based methods achieve higher reconstruction quality, the reconstruction process still suffers from longer running time due to the expense of heavy computational burden for dictionary training and sparse coding.

Fueled by the success of deep learning [15], many techniques have been proposed to implement efficient and accurate MRI reconstruction. These approaches mainly can be categorized as follows: supervised and unsupervised scheme [16]. Supervised learning is a learning technique that uses pairs of data samples [2], [17]-[24]. For instance, Schlemper *et al.* trained a deep cascade of convolutional neural networks (DC-CNN) to map undersampled data to ground-truth data [17]. Aggarwal *et al.* introduced a model-inspired (MoDL) network learning framework with a convolution neural network (CNN) based regularization prior [19]. Liu *et al.* unrolled the iterative feature refinement procedures in IFR-CS to a supervised model-driven network, dubbed IFR-Net [24]. However, it is revealed that the training of supervised learning requires a huge number of labeled samples and a great deal of expertise to ensure proper convergence. Generative adversarial networks (GAN) has shown great potential in solving CS-MRI [25]-[27]. Although it is originally used as an unsupervised learning way, a number of GAN-based algorithms fall into the supervised scheme, as they only combine the popular end-to-end network with discriminator of GAN. For instance, Conditional generative adversarial network-based approach (DAGAN) [2] not only designed end-to-end U-Net network to reduce aliasing artifacts, but also coupled the adversarial loss with an innovative content loss. Both image and frequency domains are incorporated in the network.

Unsupervised learning systems are used when paired data labels are not available. Autoencoder (AE) is a type of neural network for learning efficient data encoding in unsupervised way. Until now, two variants of AE were applied to MRI reconstruction. Particularly, Tezcan *et al.* used variational autoencoders (VAE) as a tool for describing the data density prior (DDP) [28]. Liu *et al.* [29] employed enhanced denoising autoencoders (DAE) as prior to MRI reconstruction (EDAEPRec), which leveraged DAE network as an explicit prior and maximized the likelihood by backpropagating the autoencoder error for MRI reconstruction. Zhang *et al.* [30]

This work was supported in part by the National Natural Science Foundation of China under 61871206, 61661031 and project of innovative special funds for graduate students in Jiangxi province (YC2017-S108, CX2019082). C. Quan and J. Zhou are co-first authors. (*Corresponding authors: Dong Liang, Qiegen Liu.*)

C. Quan, J. Zhou, Y. Zhu and Q. Liu are with the Department of Electronic Information Engineering, Nanchang University, Nanchang 330031, China. ({quancong, zhoujinjie, zhuyuanzheng}@email.ncu.edu.cn, liuqiegen@ncu.edu.cn).

Y. Chen is with the Key Laboratory of Computer Network and Information Integration (Southeast University), Ministry of Education, Nanjing, China (chenyang.list@seu.edu.cn)

S. Wang and D. Liang are with Paul C. Lauterbur Research Center for Biomedical Imaging, SIAT, Chinese Academy of Sciences, Shenzhen 518055, China (sophiasswang@hotmail.com, dong.liang@siat.ac.cn).

further exploited the gradient of the data density prior to reconstruction, termed EDMSPRec. Though these two works provided promising results, unlike VAE in [28], they are not directly related to generative models, lacking intuitive interpretation in machine learning terminology. Besides of the variants of AE, auto-regressive generative models like the PixelCNN were verified in MRI reconstruction recently [31].

Lately, some underlying theoretic schemes with regard to DAE were reported by different research groups. Vincent *et al.* proposed denoising score matching (DSM) and revealed that DAE is equivalent to performing DSM [32], [33]. Song *et al.* introduced a new generative model named noise conditional score networks (NCSN), where samples are produced via Langevin dynamics using gradients of the data distribution estimated with DSM progressively [34]. Motivated by the aforementioned relations that DAE and DSM can be used as good tools for generative model [35], [36], in this study we leverage this tool to accelerate the MRI reconstruction. Specifically, we explicitly immerse in embedded priors on gradients of the auxiliary data distribution to attain higher-dimensional gradients, to recover MR images more precisely. Essentially, constructing and conducting noisy prior in higher-dimensional space favors to tackle the low-dimensional manifold and low data density region issues in generative models. Results indicated that our proposal standed out for its efficiency. We highlight our main contributions as follows:

- By investigating the two major issues of low-dimensional manifold hypothesis and low data density region that occurred in estimating gradients of the data distribution, an enhanced method equipped with a higher-dimensional structure is proposed.
- Rather than the classical iterative algorithms that integrating prior information in iterative reconstruction, our algorithm views the observation as a conditional variable and incorporates it into the homotopic generative procedure.

The rest of this paper is presented as follows. Section II briefly describes some relevant works on CS-MRI and the relationship between DAE and DSM. In Section III, we elaborate the formulation of HGGDP at the prior training and iterative reconstruction stages based on higher-dimensional embedding and hotomopy techniques, respectively. Section IV presents the MRI reconstruction performance of the proposed scheme, including the experimental setup, reconstruction comparisons and ablation study. Section V concludes with topics and future works are given for the end. To facilitate the description of the present work, we tabulate the notations used hereafter in Table I.

TABLE I SUMMARIZATION OF NOTATIONS

| Notation | | Notation | |
|---|---|---|---|
| $y$ | Undersampled k-space | $p_{data}(x)$ | Data distribution |
| $x$ | Image to be reconstructed | $\tilde{x}$ | Noise-corrupted version of $x$ |
| $p$ | Undersampling matrix | $d$ | Space dimension |
| $F$ | Full Fourier encoding matrix | $k$ | K-space location |
| $F_p$ | Partially encoding matrix | $H(x)$ | High-dimensional transform |
| $x_0$ | Zero-filled reconstruction | $g_\sigma(\eta)$ | Gaussian kernel |
| $\eta$ | Artificial Gaussian noise | $p_\sigma(x)$ | Data distribution at $\sigma$-scale |
| $\sigma$ | Standard deviation | $S$ | Score network |
| $A_\sigma$ | DAE network function | $\theta$ | Parameter in the network |
| $E$ | Expectation value | $z$ | Variable in normal distribution |
| $T$ | Number of inner loop | $\alpha$ | Step size |

## II. PRELIMINARIES

### A. CS-MRI

In CS theory [4], assuming $x$ is an MR image to be reconstructed and $y$ is the under-sampled k-space data, the reconstructed image can be estimated by solving the following optimization:

$$\underset{x}{Min} \ \|F_p x - y\|^2 + \lambda R(x) \tag{1}$$

where $F_p = PF$ is the measurement matrix, $P$ is a under-sampling matrix and $F$ is a Fourier transform. Parameter $\lambda$ balances the importance of data-fidelity term $\|F_p x - y\|^2$ and regularization term $R(x)$.

Classical model-based CS-MRI approaches often adopt an iterative scheme to alternatively update the intermediate solution constrained by data-consistency and prior information. Early supervised deep learning-based CS-MRI [37], [38] learns a mapping function $x = f(x_0)$ between the zero-filling reconstructed image $x_0$ and fully-sampled reconstructed image $x$ by training a CNN with lots of paired data. The supervised end-to-end learning approaches essentially use a discriminative fashion to learn an implicit prior, lacking flexibility and robustness. In this work, we turn to the explicit prior construction for CS-MRI via generative models.

### B. From DAE to DSM

By taking advantage of the high nonlinearity and capacity of neural networks, AE has become one of the most important unsupervised learning methods for the task of representation learning [17], [39]. In order to improve the insensitivity of AE on input data, DAE tackles a partially corrupted input and is trained to recover the original undistorted input [13], [28], [29],[39]. Essentially, denoising in DAE is acted as a regularized criterion for learning to capture useful structure from the input data.

Bigdeli *et al.* [40] used the magnitude of the autoencoder error as a prior (DAEP) for image recovery and the DAE $A_\sigma$ is trained to minimize the expectation over all input images $x$:

$$L_{DAE}(A) = E_{x,\eta}[\|A_\sigma(x+\eta) - x\|^2] \tag{2}$$

where the output $A_\sigma(x)$ is trained by adding artificial Gaussian noise $\eta$ with standard deviation $\sigma$. Due to this observation, experiments by Liu *et al.* [29] shown that incorporating DAE prior into highly undersampling MRI reconstruction can achieve promising performance.

According to [40], the network output $A_\sigma(x)$ is related to the true data density $p(x)$ as follows:

$$\begin{aligned} A_\sigma(x) &= x - \frac{\int g_\sigma(\eta) p(x-\eta) \eta d\eta}{\int g_\sigma(\eta) p(x-\eta) d\eta} \\ &= x + \sigma^2 \nabla \log \int g_\sigma(\eta) p(x-\eta) d\eta \end{aligned} \tag{3}$$

where $g_\sigma(\eta)$ represents a local Gaussian kernel with standard deviation $\sigma$. DAE learns a mean-shift vector field from a given set of data samples, which is proportional to the gradient of the logarithm of the prior. Hence, a new prior called deep mean-shift prior (DMSP) was proposed by Bigdeli *et al.* [41]. They utilized it in a gradient descent approach to perform Bayes risk minimization, i.e.,

$$\nabla prior(x) = \nabla \log \int g_\sigma(\eta) p(x+\eta) d\eta \quad (4)$$
$$= [(A_\sigma(x) - x)]/\sigma^2$$

By extending the naive DMSP with integration of multi-model aggregation and multi-channel network learning, Zhang et al. [30] proposed a high-dimensional embedding network derived prior, and applied the learned prior to single-channel MRI reconstruction via augmentation technique.

Recently, Vincent showed that there is an equivalence between DAE and DSM [33]. Furthermore, Block et al. shed new light on what the DAE learns from a distribution, revealing that optimizing DAE loss is equivalent to optimizing DSM loss [42]:

**Theorem 1** [42]. The DAE loss
$$L_{DAE}(A) = E_{x \sim p, \eta \sim g_\sigma}[\|A(x+\eta) - x\|^2] \quad (5)$$
and the DSM loss
$$L_{DSM}(S) = E_{p_\sigma}[\|S(x) - \nabla \log p_\sigma(x)\|^2] \quad (6)$$
with
$$S(x) = \frac{A(x) - x}{\sigma^2} \quad (7)$$
are equivalent up to a term that does not depend on $A$ or $S$.

More details of **Theorem 1** are provided in **Supplementary material Supp-1**. As seen from DAE in **Theorem 1**, DSM optimizes the data distribution more directly. In this work, based on the connection between DAE and DSM, we will extend our preliminary idea of multi-channel and multi-view noise schemes [29], [30] on DAE to the minimization on gradients of the data distribution. More importantly, we will reveal the underlying benefits of this idea from the viewpoint of machine learning.

### C. Analysis of NCSN

Generative models are well studied and have shown outstanding performance in modeling data distributions to capitalize on known input invariances that might form good prior knowledge for informing the parameter learning [45], [46]. However, most existing generative models, either optimizing log-likelihood or adversarial training, are devoted to describing the data distribution faithfully. For instance, VAE [47] provides a formulation and a sampling approach to generate more details of images, but the samples often suffer from being blurry.

In [34], Song et al. presented NCSN that trains a score network $S_\theta(x, \sigma)$ directly to estimate the gradient of data density $\nabla_x \log p_\sigma(x)$, instead of data density $p_\sigma(x)$, and employed annealed Langevin dynamics for image generation (i.e., $p_\sigma(x) \approx p_{data}(x)$, $\sigma \to 0$). More precisely, NCSN estimates the gradients of the data distribution with DSM. Empirically, DSM is adopted as it is slightly fast and naturally fits the task of estimating scores of noise-perturbed data distributions. i.e., Assuming the noise distribution is chosen to be $p_\sigma(\tilde{x}|x) = N(\tilde{x}|x, \sigma^2 I)$, subsequently it leads to $\nabla_{\tilde{x}} \log p_\sigma(\tilde{x}|x) = -(\tilde{x}-x)/\sigma^2$. For a given $\sigma$, the objective of DSM is

$$\ell(\theta; \sigma) \triangleq \frac{1}{2} E_{p_\sigma(x)}[\|S_\theta(x,\sigma) - \nabla_x \log p_\sigma(x)\|_2^2]$$
$$= \frac{1}{2} E_{p_\sigma(\tilde{x},x)}[\|S_\theta(\tilde{x},\sigma) - \nabla_{\tilde{x}} \log p_\sigma(\tilde{x}|x)\|_2^2] + C \quad (8)$$
$$= \frac{1}{2} E_{p_\sigma(\tilde{x},x)}[\|S_\theta(\tilde{x},\sigma) + (\tilde{x}-x)/\sigma^2\|_2^2] + C$$

where $C$ is a constant that does not depend on $\theta$. Detailed statistical derivation to Eq. (8) and a new intuitive derivation are provide in **Supplementary material Supp-2 and 3**.

After Eq. (8) is derived, the DSM loss is combined for all $\sigma \in \{\sigma_i\}_{i=1}^I$ to get one unified objective

$$L(\theta; \{\sigma_i\}_{i=1}^I) \stackrel{\Delta}{=} \frac{1}{I} \sum_{i=1}^I \lambda(\sigma_i) \ell(\theta; \sigma_i) \quad (9)$$

where $\lambda(\sigma_i) > 0$ is a coefficient function depending on $\sigma_i$. Since Eq. (9) is a conical combination of $I$ DSM objectives, the optimial score $S_{\theta^*}(x, \sigma)$ minimizes Eq. (9) if and only if $S_{\theta^*}(x, \sigma_i) = \nabla_x \log p_{\sigma_i}(x)$ is satifised for all $i \in \{1, 2, \cdots, I\}$.

As the training procedure of the NCSN $S_\theta(x, \sigma)$ is determined, annealed Langevin dynamics as a sampling approach is introduced. It recursively computes the following:

$$x_t = x_{t-1} + \frac{\alpha_i}{2} \nabla_x \log p_{\sigma_i}(x_{t-1}) + \sqrt{\alpha_i} z_t$$
$$= x_{t-1} + \frac{\alpha_i}{2} S_\theta(x_{t-1}, \sigma_i) + \sqrt{\alpha_i} z_t \quad (10)$$

where $\alpha_i$ is the step size by tuning down it gradually. $T$ is the number of iterations for each noise level, and $z_t \sim N(0,1)$.

To sum up, two key contributions of NCSN can be summarized as follows: One is perturbing the data with random Gaussian noise of various magnitudes; The other is that an annealed version of Langevin dynamics is initialized using scores corresponding to the highest noise level, and gradually annealing down the noise level until it is small enough to be indistinguishable from the original data distribution. Essentially, the core innovation behind the above contributions is the flexible usage of artificial noise.

Although NCSN has achieved good results, its application in MRI reconstruction is still leaving huge room for improvement, particularly on two major deficiencies: low data density regions and the manifold hypothesis [48], [49]. For instance, it is often assumed that the data distribution is supported on a low dimensional manifold. Then, the score will be undefined in the ambient space such that matching method will fail to provide a consistent score estimator. Additionally, the scarcity of training data in low data density regions, e.g., far from the manifold, hinders the matching accuracy and slows down the mixing of Langevin dynamics sampling. To circumvent these limitations, we propose HGGDP to enrich the native NCSN by the usage of artificial noise in higher-dimensional space. The underlying idea is to span the whole space with more samples to override the deficiency from the manifold hypothesis and enable the data not be confined to low data density regions.

### III. PROPOSED HGGDP MODEL

In the previous section, we discuss the DAE and DSM, where the estimation is done from data density to gradient of data density. In this section, we concentrate on effectively improving the performance of native DSM on MRI reconstruction, coined deep gradients of generative density prior (HGGDP). Specifically, we introduce a way of to avert the data confined to low data density regions and avoid difficulties from the manifold hypothesis, i.e., forming high-dimensional tensor and injecting artificial noise.

The whole process of HGGDP and HGGDPRec is shown in Fig. 1. The original MR data manifold can be seen in Fig. 1(a). Fig. 1(b) illustrates the characteristics of DSM, which learns to reconstruct the clean input from a corrupted version. Fig. 1(c) visualizes the phenomenon that DSM is not accurate for data at a relatively low-dimensional manifold [43] with single level of noise. There exists a issue for SM: The density, or the gradient of the density is undefined outside the manifold, making it difficult to train a valid density model for the data distribution defined on the entire space. Hence, how to reduce the low data density regions is urgent.

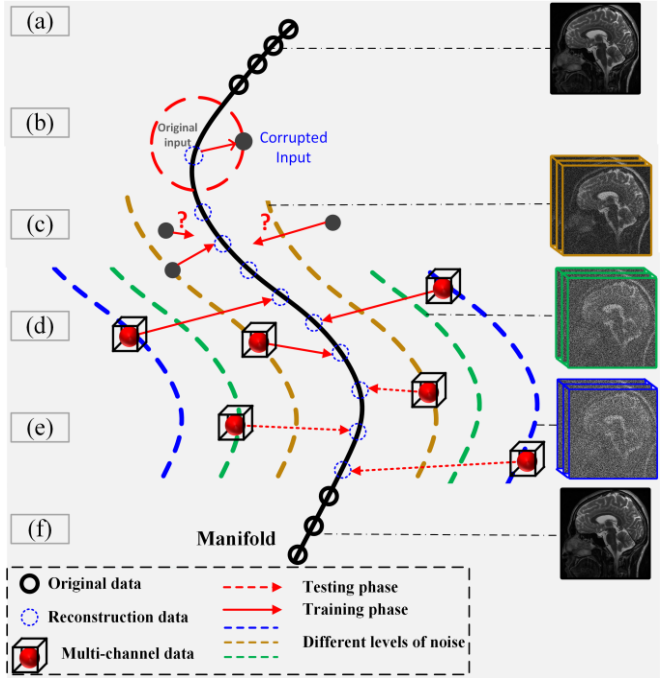

**Fig. 1.** Schematic diagram of manifold learning using HGGDP. (a) Manifold of ground-truth data. (b) Corrupted Input. (c) DSM with single noise scale. (d)(e) Training and reconstruction stage via projecting on multi-channel space with multiple noise scales. (f) Purified data. The legend is located at the bottom.

To alleviate the difficulty encountered in previous models based on DSM with single noise scale when sampled with noise initialization, prior learning and sample generation under multiple noise scales [29][34][43] was developed. To further inherit the spirit in [29][34][43], we enhance the DSM by multi-channel MR images with multi-view noise. The training and reconstruction phases are visualized in Fig. 1(d)(e). To illustrate this point, in the right region of Fig. 1 we show some instances of intermediate samples during annealed sampling process. Displayed are noisy samples that carrying different levels of noise. One can see that, as the sample point approaches the data manifold (i.e., point from blue line, to green line, and to brown line), more and more details are presented in the samples. It is an ideal situation that all the sample points in training and reconstructing phase approach to the original data manifold. Thus, the purified data manifold is obtained in Fig. 1(f) finally. More results and explanations will be provided in the following contexts.

### A. HGGDP: Prior in High-dimensional Space

In regions of low data density, the original DSM may not have enough evidence to estimate score functions accurately, due to the lack of data samples. To further enrich the efficiacy of DSM in the naïve NCSN, we introduce a high-dimensional space embedding strategy to boost the representation diversity of generative modeling, and thereafter reconstruction performance.

**Theorem 2** [42]. Let $F$ be a class of $\mathbb{R}^d$-valued functions, all of which are $M/2$-Lipchitz, bounded coordinate wise by $R > 0$, containing arbitrarily good approximations of $\nabla \log p_\sigma$ on the ball of radius $R$. Let $\sigma < \sigma_{max}$ and support we have $n$ i.i.d samples from $p_\sigma$, $x_1, \cdots, x_n$. Let

$$\hat{S} \in \arg\min_{S \in F} \frac{1}{n} \sum_{i=1}^{n} \| S(x_i) - \nabla \log p_\sigma(x_i) \| \quad (11)$$

Then with probability at least $1 - 4\delta - Cne^{-R^2/m_\sigma}$ on the randomness due to the sample,

$$E_{p_\sigma}[\| \hat{S}(x) - \nabla \log p_\sigma(x) \|^2] \leq C(MR+B)^2(\log^3 n \cdot \mathfrak{R}_n^2(F) + \beta_n d) \quad (12)$$

where $m_\sigma = (m - \sigma^2 M)/2$ and $C$ is a universal constant. Both $\mathfrak{R}_n^2(F)$ and $\beta_n$ contain the factor of $1/n$. More details can be found in **Supplementary material Supp-4**.

From **Theorem 2**, it is concluded that the representation boundary is related to the sample number $n$ and space dimension $d$. Particularly, the larger the sample number is, the less the representation error will be. This observation motivates us to boost the performance via sampling in higher-dimensional embedding space. The implementation of this work consists of two consecutive operations: Forming higher-dimensional tensor and injecting artificial noise.

**Forming higher-dimensional tensor**: To take advantage of DAE as an advanced prior, EDAEPRec proposed by Liu *et al.* [29] handles complex-valued data by concatenating the real and imaginary parts as channels, that is, the real and imaginary components are set as the network input simultaneously via replication technology. They elaborated that prior information learned from high-dimensional tensor is more effective for MRI reconstruction than the low-dimensional counterpart.

Inspired by the appealing performance of EDAEPRec over the naive DAE and the connection between DAE and DSM, we use replication technology for promoting the DSM in generative model NCSN. In detail, we utilize a general transformation $X = H(x)$ to establish a $N$-channel higher-dimensional tensor e.g., the vector variable with $N$ channels as $X = \{x, x, \cdots, x\}$. The goal of stacking to be $X$ is to obtain much more data information at high-dimensional manifold and high-density regions, thus avoiding some difficulties for both accuracy in score estimation and sampling with Langevin dynamics. Subsequently, the HGGDP is trained with $X$ in high-dimensional space as network input and the parameterized $S_\theta(X, \sigma)$ is obtained.

To show the superior performance of higher-dimensional formulation strategy, we visualize the data scores $\nabla_x \log p_{data}(x)$ and the estimated scores $S_\theta(X, \sigma)$ of HGGDP in Fig. 2 (see **Supplementary material Supp-5** for the experiemtal setup). As can be seen, after expanding the space dimension, the diversity of score matching increases. Additionally, the trouble of low data density regions is alleviated as the data samples in 3D space increases.

In our case of handling complex-valued MR image, the transformation $H(x)$ is adopted to form high dimensional tensor $X$, as shown in Fig. 3. A $N$-channel high-dimensional tensor $X = \{[x_{real1}, x_{imag1}], \cdots, [x_{realN}, x_{imagN}]\}$ is the channel concatenation of real and imaginary compo-nents. In this work $X = \{[x_{real1}, x_{imag1}], [x_{real2}, x_{imag2}], [x_{real3}, x_{imag3}]\}$ with the setting of $N = 3$.

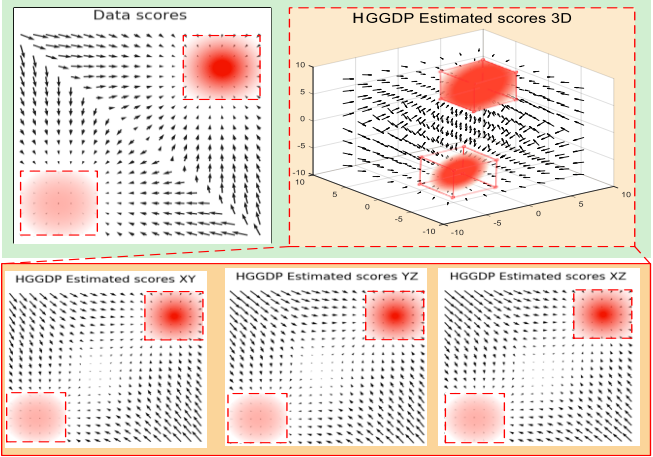

**Fig. 2**. Top: An illustration of $\nabla_x \log p_{data}(x)$ (Left) and $S_\theta(X,\sigma)$ (Right). $\nabla_x \log p_{data}(x)$ is the 2D data scores, $S_\theta(X,\sigma)$ is the 3D data scores from HGGDP. Bottom: The projection of the 3D distribution on three axes.

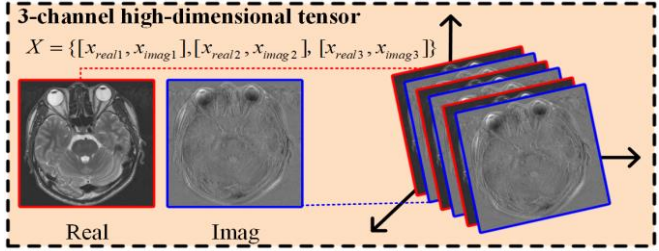

**Fig. 3**. Visualization of the three-channel high-dimensional tensor $X$.

**Injecting artificial noise**: The manifold hypothesis states that data in the real world tend to concentrate on low-dimensional manifolds embedded in an ambient space. There are two vital issues faced by score-based generative models. On one hand, when $x$ is restricted to a low dimensional manifold, the score $\nabla_x \log p_{data}(x)$ is undefined, which is a gradient taken in the ambient space. On the other hand, a consistent score estimator is afforded only when the support of the data distribution is the whole space; If the data resides on a low-dimensional manifold, it will be inconsistent. Song *et al*. discovered that perturbing the data with random Gaussian noise of various magnitudes decreased the possibility of the data distribution that collapsing to a low dimensional manifold [34].

In HGGDP, more artificial noise is injected in high-dimensional space. This is, after forming $X$ in higher-dimensional space, artificial noise injection is needed to fill in the whole space. Similarly as in Eq. (9), the unified objective is executed on all samples $X$ for all $\sigma \in \{\sigma_i\}_{i=1}^I$ to pursue the optimal score,

$$\hat{L}(\theta;\{\sigma_i\}_{i=1}^I) \triangleq \frac{1}{I}\sum_{i=1}^I \lambda(\sigma_i)\hat{\ell}(\theta;\sigma_i) \quad (13)$$

Furthermore, after learning the parameterized score of probability density $S_\theta(X,\sigma)$ from $X$, artificial noise injection is also done at the reconstruction phase of HGGDPRec. Specifically, a noisy high-dimensional tensor is produced with varying levels of Gaussian noise added to $X$. More critically, injecting noise in higher-dimensional space will produce more diverse samples in the perturbed data distribution than that in original data distribution, thus improving score estimation accuracy. Taking advantage of it, the reconstruction result is obtained by an operation that makes noisy high-dimensional tensor as the input of $S_\theta(X,\sigma)$ via gradually annealed noise. i.e.,

$$X^t = X^{t-1} + \frac{\alpha_i}{2}S_\theta(X^{t-1},\sigma_i) + \sqrt{\alpha_i}z_t \quad (14)$$

where the initial solution $X^0$ can be a total uniform noise or other pre-defined value. The index set $t = 1,\cdots,T$ is defined at the inner loop of each $\sigma_i$, while index set $i = 1,\cdots,I$ goes over the outer loop of every $\sigma_i$.

The benefit of injecting noise in high-dimensional space is visualized in Fig. 4. Fig. 4(a) shows the training losses of the objects that occurred in noisy single-channel and three-channel space, respectively. The network is trained to estimate the data score on *SIAT* dataset that perturbed with small Gaussian noise. Notice that the imposed Gaussian noise $N(0, 0.0001)$ is very small for images with pixel values in the range [0, 1], and is almost indistinguishable to human eyes. As can be seen, both loss curves first decrease and then fluctuate as the number of epochs increases. Two distinct phenomena can be observed. First, the loss with single-channel fluctuates irregularly, while the result in the three-channel circumstance is much stable, indicating better convergence property. Second, the loss value trained on three-channel MR images is much lower than the single-channel counterpart, which implies better score accuracy. In addition, Fig. 4(b) visualizes the reconstruction of NCSN and HGGDP with enlarged views and error maps. In comparison, HGGDP achieves higher reconstruction accuracy and gives more faithful result.

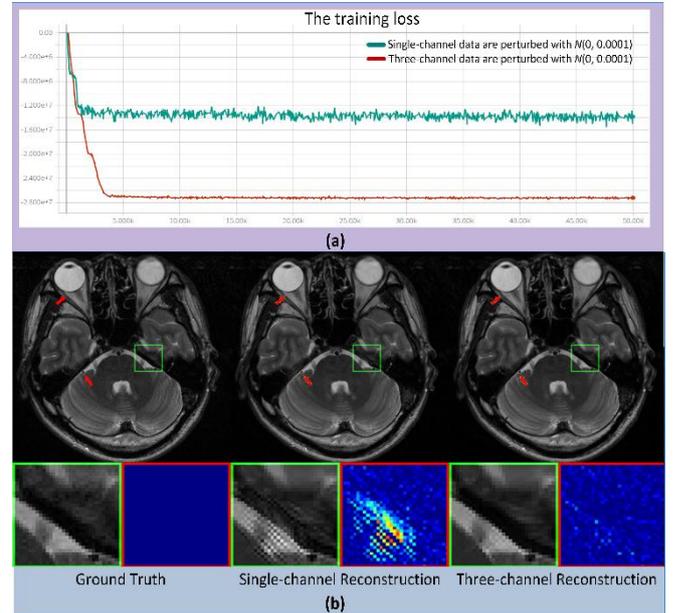

**Fig. 4**. Performance comparison of "injecting noise in high-dimensional tensor". (a) Training loss in DSM of the naïve NCSN and the advanced HGGDP at each epoch. (b) Image quality comparison on the brain data at 15% radial sampling: Reconstruction images, error maps (Red) and zoom-in results (Green).

As illustrated in Fig. 5(a), we use annealed Langevin dynamics to further sample from the high-dimensional noisy data distribution with multi-view noise. In order to intuit the procedure of annealed Langevin dynamics, we provide the intermediate samples in Fig. 5(b), where each row shows how samples evolve from pure random noise to high-resolution MR images.

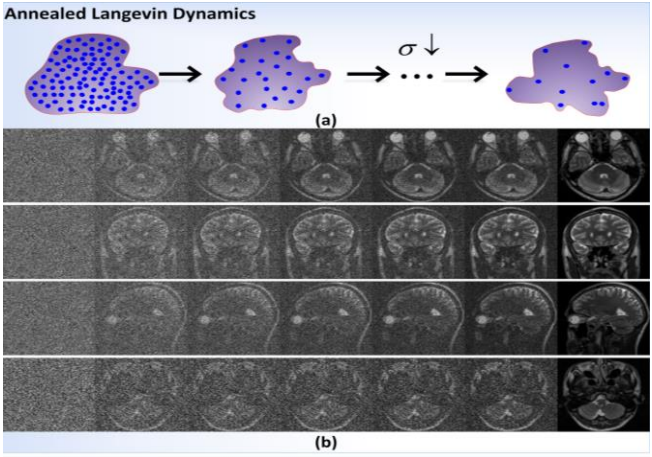

Fig. 5. Pipeline of sampling from the high-dimensional noisy data distribution with multi-view noise and intermediate samples. (a) Conceptual diagram of the sampling on high-dimensional noisy data distribution with multi-view noise. (b) Intermediate samples of annealed Langevin dynamics.

### B. HGGDPRec: Iterative Reconstruction in Homotopy Fashion

As proved in [33] and shown in Eq. (8), the estimation of DSM is optimally accurate only when $p_\sigma(x) \to p_{data}(x)$ (i.e., $\nabla \log p_\sigma(x) \to \nabla \log p_{data}(x)$), $\sigma \to 0$. Therefore, a homotopic process is needed to attain accurate density prior. Rather than the classical iterative algorithms that integrating prior information in iterative reconstruction, the proposed HGGDPRec views the observation as a conditional variable and incorporates it into the iterative generative procedure.

Specifically, the generative procedure of HGGDPRec for MRI reconstruction is a homotopic process: At the iterative stage, we generate a list of noise levels $\{\sigma_i\}_{i=1}^I$ that are reduced proportionally for each step of the outer loop. At each stage, we add artificial noise to the intermediate solution according to the noise level $\sigma_i$ from large to small at the iterative reconstruction. The transition helps smoothly transfer the benefits of large noise levels to low noise levels where the perturbed data are almost indistinguishable from the original one. In the meantime, the data consistency of the partial K-space measurement is incorporated into the update of iterative generative process, forming the iterative reconstruction procedure. Finally, at $\sigma_I$, the influence of artificial noise is no longer recognized, HGGDPRec has reached stable and perfect reconstruction quality.

It should be emphasized that the annealed/homotopic strategy used in HGGDPRec has a long history in severely non-convex optimization for reconstruction. For instance, in [50], Trzasko et al. aimed to attain highly undersampled MRI reconstruction via homotopic L0-minimization. In [51], Wong et al. extended the homotopic L0-minimization scheme for regional sparsified domain. In those works, they continually reduce the parameter in the regularization term to pursue the global solution in non-convex minimization. Here, we adopt a similar strategy in the case of statistical distribution learning to minimize the non-smooth and non-convex functional $\nabla_X \log p_{data}(X)$.

Specifically, at each iteration of the annealed Langevin dynamics, we update the solution via data consistency constraint after Eq. (14), i.e., let $x^t = Mean(X^t)$, it yields

$$x^{t+1} = \arg\min_x \|y - F_p x\|^2 + \lambda \|x - x^t\|^2 \quad (15)$$

The least-square (LS) minimization in Eq. (15) can be solved as follows:

$$(F_p^T F_p + \lambda) x^{t+1} = F_p^T y + \lambda x^t \quad (16)$$

Let $F \in C^{M \times M}$ denotes the full Fourier encoding matrix which is normalized as $F^T F = 1_M$. $Fx(k_v)$ stands for the updated value at under-sampled k-space location $k_v$, and $\Omega$ represents the sampled subset of data, it yields,

$$Fx(k_v) = \begin{cases} Fx^t(k_v), & k_v \notin \Omega \\ \dfrac{FF_p^T y(k_v) + \lambda Fx^t(k_v)}{(1+\lambda)}, & k_v \in \Omega \end{cases} \quad (17)$$

In summary, as explained in **Algorithm 1**, the whole MRI reconstruction procedure consists of two-level loops: The outer loop handles $\nabla_X \log p_\sigma(X)$ to approximate $\nabla_X \log p_{data}(X)$, while the inner loop decouples to be an alternating process of the estimated gradient of data prior $\nabla_X \log p_\sigma(X)$ and the LS scheme. For the algorithm convergence, both the iteration of Langevin dynamics [42] and the homotopy strategy with finite states [50] have convergence guarantee. By incorporating the analytic solution in Eq. (17) into the iterative procedure, the overall HGGDPRec algorithm will tend to the convergence zone after finite iterations. Detailed flowchart of HGGDPRec is shown in Fig. 6.

| Algorithm 1 HGGDPRec |
|---|
| **Training stage** |
| **Dataset**: Multi-channel dataset: $X = \{X_1, X_2, X_3, \cdots X_N\}$ |
| **Output**: Trained HGGDP $S_\theta(X, \sigma)$ |
| **Reconstruction stage** |
| **Setting**: $\sigma \in \{\sigma_i\}_{i=1}^I$, $\varepsilon$, $T$, $x^0$, $k_v$ and $\Omega$ |
| 1: **for** $i \leftarrow 1$ to $I$ **do (Outer loop)** |
| 2:     $\alpha_i = \varepsilon \cdot \sigma_i^2 / \sigma_I^2$ |
| 3:     **for** $t \leftarrow 1$ to $T$ **do (Inner loop)** |
| 4:         Draw $z_t \sim N(0,1)$ and $X^{t-1} = \{x^{t-1}, x^{t-1}, \cdots, x^{t-1}\}$ |
| 5:         $X^t = X^{t-1} + \dfrac{\alpha_i}{2} S_\theta(X^{t-1}, \sigma_i) + \sqrt{\alpha_i} z_t$ |
| 6:         Update $x^t = Mean(X^t)$ and Eq. (17) |
| 7:     **end for** |
| 8:     $x^0 \leftarrow x^T$ |
| 9: **end for** |
| **Return** $x^T$ |

## IV. EXPERIMENTS

### A. Experiment Setup

1) **Datasets**: The experiments were performed on brain images from *SIAT*, knee images from *FastMRI*, two multi-coil acquisition data Test1 [19] and Test2 [28].

First, we use brain images from *SIAT* dataset, which was provided by Shenzhen Institutes of Advanced Technology, the Chinese Academy of Science. Informed consents were obtained from the imaging subject in compliance with the institutional review board policy. The raw data were acquired from 3D turbo spin-echo (TSE) sequence with T2 weighting by a 3.0T whole-body MR system (SIEMENS MAGNETOM Trio Tim), which has 192 slices per slab, and the thickness of each slice was 0.86 mm. Typically, the field of view and voxel size were set to be 220×220 mm$^2$ and 0.9×0.9×0.9 mm$^3$, respectively. Moreover, the number of coils is 12 and the collected dataset includes 500 2D complex-valued MR images. Affine transformation was adopted for data augmentation, and 8000 patches are obtained for training by slicing the enhanced image into 64×64. At the reconstruction stage, we use a variety of sampling schemes on another 31 complex-valued MR images.

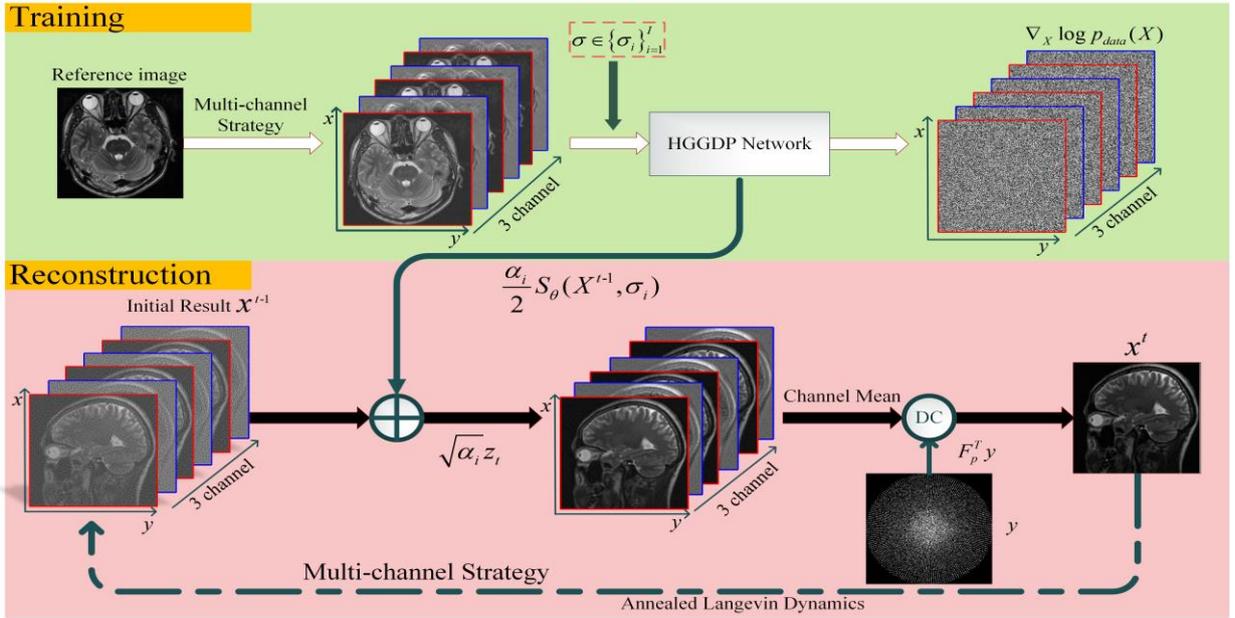

**Fig. 6**. Pipeline of HGGDP training process for prior learning and HGGDPRec procedure for MRI reconstruction.

Second, experiments were conducted on single coil knee images from *FastMRI* dataset. The following sequence parameters were used: Echo train length 4, matrix size 320 × 320, in-plane resolution 0.5mm×0.5mm, slice thickness 3mm, no gap between slices. Timing varied between systems, with repetition time (TR) ranging between 2200 and 3000 milliseconds, and echo time (TE) between 27 and 34 milliseconds. We randomly selected 80 files (about 2941 images) and 2 files (about 50 images) from *FastMRI* to train and test our model, respectively. It is worth noting that the training set is generated in the same way as *SIAT*, which produced 26469 patches.

Finally, besides of the above two single coil datasets, experiments were done on multi-coil data Test1 [19] and Test2 [28] to verify the performance of HGGDPRec for parallel imaging reconstruction.

2) *Model Training:* During the learning phase, we use fully sampled MR images as the network input and disturb it simultaneously via random Gaussian noise of various amplitudes. Notice that the input and output are all complex-valued images with the same size and each includes real and imaginary components. Additionally, RefineNet [39] with instance normalization, dilated convolutions and U-Net-type architectures[52] were selected as the network structure. Adam was selected as an optimizer with the learning rate 0.005, which was halved every 5000 iterations. Subsequently, the HGGDP model was trained after $1e5$ iterations with the batch size of 32 that took around 20 hours. It was performed with Pytorch interface on 2 NVIDIA Titan XP GPUs, 12GB RAM.

3) *Sampling Masks*: The sampling in the Fourier domain is realized using three different undersampling strategies, namely 2D random, pseudo radial and 1D cartesian sampling. Additionally, the accelerated factors are varied over nine values, $R = 2, 3, 3.3, 4, 5, 6, 6.7, 8, 10$. Some sampling masks are depicted in Fig. 7.

4) *Compared Methods*: Several state-of-the-art methods were compared against HGGDPRec. For single coil, we implemented patch-based DLMRI [53], reference-derived sparse representation method PANO [54], dictionary learning method FDLCP [55], low-rank-based NLR-CS [56], end-to-end DC-CNN [17], enhanced denoising autoencoder based EDAEPRec [29], neural proximal gradient descent (NPGD) [58], and NCSNRec [34] on *SIAT*. We additionally ran low-rank modeling of local k-space neighborhoods LORAKS [59], supervised learning U-Net and DC-CNN [17], and multi-channel enhanced deep mean-shift prior MEDMSPRec [30] on *FastMRI*. For parallel imaging, we implemented VAE-based DDP [19] and model-based MoDL [28] on two multi-coil brain images, respectively.

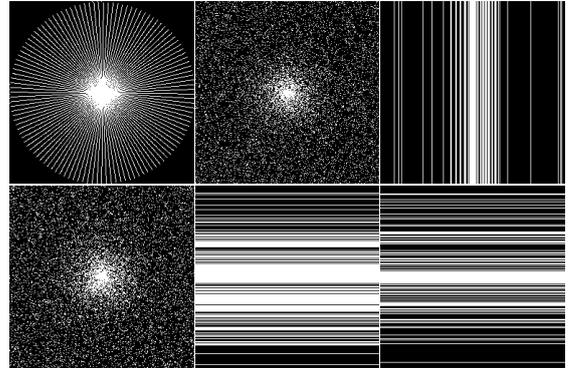

**Fig. 7**. Representative sampling masks.

Additionally, we quantified the reconstruction performance by peak signal-to-noise ratio (PSNR), structural similarity (SSIM) and high-frequency error norm (HFEN). For convenient reproducibility, source code is available at: https://github.com/yqx7150/HGGDPRec.

B. *Single Coil Reconstruction Comparisons*

To evaluate the reconstruction performance of HGG-DPRec in the case of single coil reconstruction, 31 complex-valued MR images from *SIAT* and 50 proton-density weighted knee images from the "single coil train" set of *FastMRI* are used in this subsection.

Table II summarizes the quantitative PSNR/SSIM/HFEN results of *SIAT* with various sampling rates and patterns. Intuitively, the average results of 31 test images obtained by HGGDPRec are more than 2 dB over the naïve NCSNRec. Furthermore, HGGDPRec yields the highest values in the most majority of the sampling rates. For example, under the accelerated factors $R = 4, 5$, the highest PSNR values achieved by all the compared methods are 34.49 dB and

33.49 dB, which are lower than the values of 35.55 dB and 34.40 dB obtained by HGGDPRec. Even though NPGD shows appealing reconstruction results at low under-sampling rates, HGGDPRec is more effective at high under-sampling rates. Additionally, we evaluate the computational costs of HGGDPRec and the competing methods in Table III. As can be seen, with GPU implementation, the computational cost of HGGDPRec is moderate.

Besides of the quantitative comparison, the visual quality is also highlighted. As can be seen in Figs. 8-9, visual quality of reconstructions for different methods varies. DLMRI and NLR-CS reconstructions have limitations in recovering the structure and texture. FDLCP utilizes the similarity and geometric orientation of the patch, which contains more details than DLMRI. In order to further compare visual quality, the zoom-in images and error maps are illustrated through the green and red boxes on the screen. The visual quality indicates that EDAEPRec is better than FDLCP, PANO and DC-CNN methods in terms of more abundant recovered edge details. However, it still suffers from some undesirable artifacts and loses details such as ringing, jaggy and staircase artifacts. To sum up, HGGDPRec can achieve more satisfactory results with clearer contours, sharper edges, and finer image details under various sampling masks. The convergence tendency of PSNR and HFEN curves versus iteration for reconstructing the 12th brain image is plotted in Fig. 10. It can be seen that these two curves are wavy at early iterations and then become stable.

The entire reconstruction process is convergent. It is also concluded that the total cost time can be reduced by reducing iteration number of the inner loop.

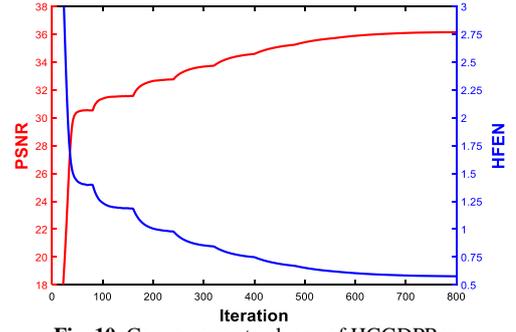

**Fig. 10**. Convergence tendency of HGGDPRec.

Additionally, comparisons to LORAKS, U-Net, DC-CNN and MEDMSPRec are conducted in this experiment. Specifically, LORAKS is a flexible framework for constrained image reconstruction that uses low-rank matrix modeling of local k-space neighborhoods. The supervised learning U-Net and DC-CNN are trained and validated by 977 files (about 37216 images) and 200 files (about 7600 images) randomly selected from "single coil train" set of *FastMRI*, respectively. MEDMSPRec exploits prior from the multi-channel MWCNN, whose network architecture is similar to U-Net, it attains better performance.

TABLE II
AVERAGE PSNR, SSIM AND HFEN VALUES OF RECONSTRUCTING 31 TEST IMAGES BY DIFFERENT ALGORITHMS AT RADIAL SAMPLING TRAJECTORIES AND DIFFERENT SAMPLING TRAJECTORIES WITH THE SAME PERCENTAGE IN SINGLE COIL.

| (a) | | DLMRI | PANO | FDLCP | NLR-CS | DC-CNN | RefineGAN | MEDSMPRec | EDAEPRec | NPGD | NCSNRec | HGGDPRec |
|---|---|---|---|---|---|---|---|---|---|---|---|---|
| $R$=3.3, Pseudo Radial | PSNR | 33.43 | 34.64 | 34.89 | 35.31 | 35.71 | 34.91 | 34.76 | 35.62 | **36.98** | 34.08 | 36.18 |
| | SSIM | 0.9054 | 0.9152 | 0.9135 | 0.9099 | 0.9234 | 0.9380 | 0.9193 | 0.9279 | **0.9653** | 0.9090 | 0.9364 |
| | HFEN | 0.63 | 0.56 | 0.50 | 0.47 | 0.44 | 0.47 | 0.58 | 0.42 | **0.41** | 0.76 | 0.42 |
| $R$=4, Pseudo Radial | PSNR | 32.41 | 33.65 | 34.04 | 34.35 | 34.07 | 32.81 | 34.11 | 34.49 | **36.17** | 33.51 | 35.55 |
| | SSIM | 0.8866 | 0.8995 | 0.8980 | 0.8938 | 0.8992 | 0.9188 | 0.9098 | 0.9151 | **0.9578** | 0.9013 | 0.9288 |
| | HFEN | 0.84 | 0.73 | 0.62 | 0.61 | 0.69 | 0.73 | 0.68 | 0.64 | **0.50** | 0.85 | **0.50** |
| $R$=5, Pseudo Radial | PSNR | 31.21 | 32.44 | 32.97 | 33.32 | 32.68 | 30.52 | 33.00 | 33.49 | 34.01 | 32.59 | **34.40** |
| | SSIM | 0.8602 | 0.8777 | 0.8770 | 0.8812 | 0.8791 | 0.8718 | 0.8908 | 0.8990 | **0.9385** | 0.8868 | 0.9130 |
| | HFEN | 1.10 | 0.96 | 0.80 | 0.79 | 0.95 | 1.19 | 0.86 | 0.79 | 0.73 | 1.01 | **0.65** |
| $R$=10, Pseudo Radial | PSNR | 27.39 | 28.58 | 29.34 | 29.51 | 28.39 | 25.84 | 29.91 | 30.30 | 29.95 | 29.76 | **30.88** |
| | SSIM | 0.1444 | 0.7805 | 0.7856 | 0.7845 | 0.7710 | 0.7311 | 0.8206 | 0.8319 | **0.8593** | 0.8330 | 0.8488 |
| | HFEN | 2.18 | 1.90 | 1.60 | 1.65 | 1.93 | 2.36 | 1.48 | 1.40 | 1.62 | 1.57 | **1.29** |
| (b) | | DLMRI | PANO | FDLCP | NLR-CS | DC-CNN | RefineGAN | MEDSMPRec | EDAEPRec | NPGD | NCSNRec | HGGDPRec |
| $R$=6.7, 2D Random | PSNR | 27.63 | 29.12 | 30.14 | 30.34 | 28.78 | 25.95 | 30.36 | 30.68 | 30.57 | 30.18 | **31.78** |
| | SSIM | 0.7518 | 0.7964 | 0.8004 | 0.8087 | 0.7873 | 0.7205 | 0.8350 | 0.8433 | **0.8671** | 0.8445 | 0.8649 |
| | HFEN | 2.02 | 1.77 | 1.44 | 1.46 | 1.83 | 2.34 | 1.38 | 1.31 | 1.52 | 1.47 | **1.16** |
| $R$=6.7, Pseudo Radial | PSNR | 29.36 | 30.60 | 31.31 | 31.35 | 30.57 | 27.94 | 31.54 | 32.00 | 32.48 | 31.3185 | **32.79** |
| | SSIM | 0.7518 | 0.8372 | 0.8391 | 0.8494 | 0.8348 | 0.8145 | 0.8616 | 0.8716 | **0.9125** | 0.8658 | 0.8873 |
| | HFEN | 2.02 | 1.37 | 1.13 | 1.17 | 1.38 | 1.70 | 1.13 | 1.05 | 1.07 | 1.24 | **0.91** |
| $R$=6.7, 1D Cartesian | PSNR | 26.50 | 27.51 | 27.91 | 28.23 | 27.05 | 26.05 | 28.45 | 28.85 | 28.64 | 28.53 | **29.05** |
| | SSIM | 0.7390 | 0.7683 | 0.7776 | 0.7798 | 0.7506 | 0.7771 | 0.7935 | 0.8041 | **0.8462** | 0.8175 | 0.8225 |
| | HFEN | 2.51 | 2.28 | 2.15 | 2.03 | 2.44 | 2.42 | 1.91 | 1.81 | 2.08 | 1.92 | **1.78** |

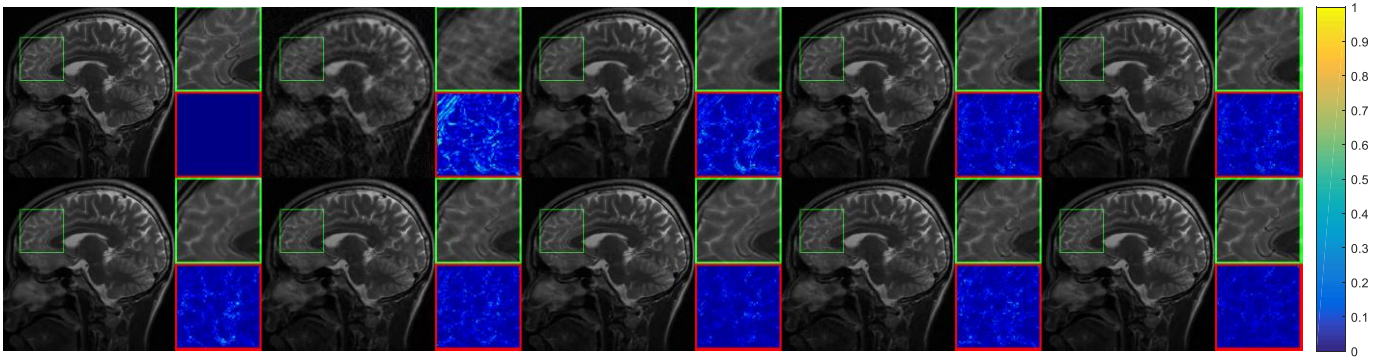

**Fig. 8**. Reconstruction comparison on pseudo radial sampling at acceleration factor $R = 5$ in single coil. Top: Reference, reconstruction by Zero-filled, DLMRI, PANO, FDLCP; Bottom: Reconstruction of NLR-CS, DC-CNN, EDAEPRec, NPGD, HGGDPRec. Green and red boxes illustrate the zoom in results and error maps, respectively.

TABLE III
THE ORIGINAL IMAGE IS A 256×256 MRI IMAGE. NOTICE THAT DLMRI AND NLR-CS ARE EXECUTED WITH CPU, WHILE PANO, FDLCP, DC-CNN, RE-FINEGAN, MEDMSPREC, EDAEPREC, NPGD, NCSNREC AND HGGDPREC ARE WITH GPU.

| Runtime(s) | DLMRI | PANO | FDLCP | NLR-CS | DC-CNN | Refi-neGAN | MED-MSPRec | EDAEPRec | NPGD | NCSNRec | HGG-DPRec |
|---|---|---|---|---|---|---|---|---|---|---|---|
| Total time | 430.36 | 13.74 | 83.28 | 127.32 | 13.51 | 9.10 | 131.42 | 82.88 | 0.14 | 230.72 | 244.86 |
| Iter time | 17.21/iter | 6.87/iter | 41.64/iter | 0.91/iter | --/-- | --/-- | 0.44/iter | 0.42/iter | --/-- | 0.21/iter | 0.26/iter |

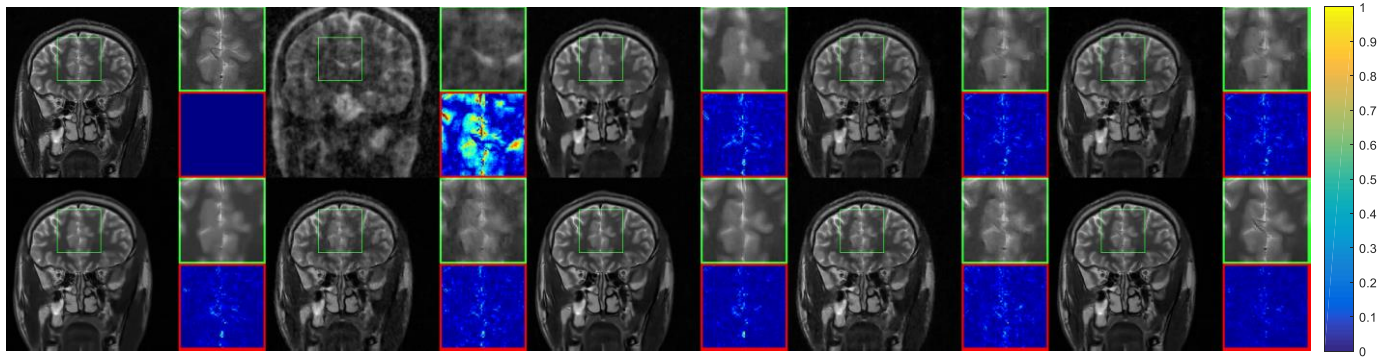

**Fig. 9**. Reconstruction comparison on 2D Random sampling at acceleration factor $R=6.7$ in single coil. Top: Reference, reconstruction by Zero-filled, DLMRI, PANO, FDLCP; Bottom: Reconstruction of NLR-CS, DC-CNN, EDAEPRec, NPGD, HGGDPRec. Green and red boxes illustrate the zoom in results and error maps, respectively.

Table IV reports the average results of 50 reconstructed images. As can be observed, HGGDPRec outperforms LORAKS, U-Net, DC-CNN and MEDMSPRec significantly. Theoretically, the inferiority of LORAKS indicates that the traditional model is insufficient to tackle with the challenging data with amounts of high-resolution features and details. Although the supervised deep learning approach DC-CNN works much better than U-Net, it still has large room to improve. Both the unsupervised schemes MEDMSPRec and HGGDPRec exploit prior from higher-dimensional space, however, the former only estimate the approximate gradient of data prior $\nabla_x \log p_\sigma(x)$ and the latter tries to learn the gradient of the true data prior $\nabla_x \log p_{data}(x)$. Compared with the competing algorithms, HGGDPRec achieves better quantitative results under both 1D Cartesian samplings at $R=4$ and $R=8$ in single coil, in terms of lower HFEN, higher PSNR and SSIM values.

Fig. 11 depicts the qualitative comparison results with two reconstruction slices at 1D Cartesian sampling under acceleration factor $R=4$ and $R=8$, respectively. As indicated, the reconstructions with Zero-filled at a high acceleration rate create significant aliasing artifacts and lose anatomical details. LORAKS can improve the reconstruction as compared to Zero-filled image, but it is hard to see a significant improvement when a amount of aliasing artifact is presented. U-Net and DC-CNN have excellent reconstruction effects in some areas, but the overall reconstruction quality is still lower than HGGDPRec. To sum up, the proposed HGGDPRec provides realistic image quality and preserves the detailed structures as well as textures.

TABLE IV
AVERAGE PSNR, HFEN AND SSIM RESULTS OF FIVE METHODS ON 50 COMPLEX-VALUED KNEE IMAGES IN SINGLE COIL.

| Method | $R=4$ | | | $R=8$ | | |
|---|---|---|---|---|---|---|
|  | PSNR | SSIM | HFEN | PSNR | SSIM | HFEN |
| LORAKS | 28.49 | 0.7075 | 1.72 | 26.05 | 0.6371 | 2.15 |
| U-Net | 30.05 | 0.7894 | 1.45 | 26.92 | **0.7166** | 1.92 |
| DC-CNN | 31.37 | 0.7707 | 1.22 | 29.15 | 0.6841 | 1.68 |
| MEDMSPRec | 30.38 | 0.7652 | 1.28 | 27.97 | 0.6794 | 1.75 |
| HGGDPRec | **31.69** | **0.7939** | **0.98** | **29.38** | 0.7065 | **1.52** |

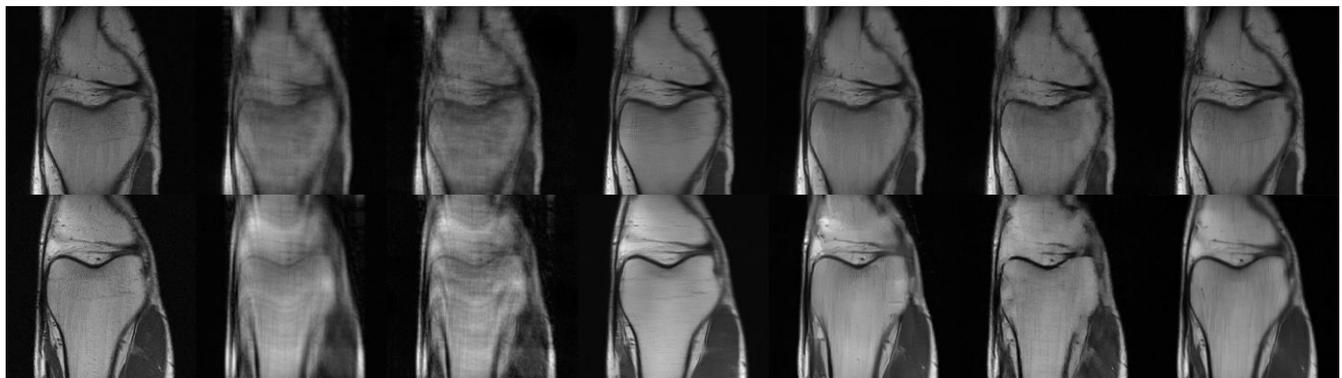

**Fig. 11.** Complex-valued reconstruction results on knee images at various 1D Cartesian undersampling patterns in single coil. From top to bottom: The reconstruction results on knee images at $R = 4$ and $R = 8$, respectively. From left to right: Ground-truth, reconstruction of Zero-filled, LORAKS, U-Net, MEDMSPRec, DC-CNN and HGGDPRec.

### C. Parallel Imaging Reconstruction Comparisons

As well known, flexibility and robustness are the main characteristic of unsupervised learning strategy that differs from the supervised learning counterpart. In this subsection, HGGDPRec is compared with MoDL and DDP for parallel imaging reconstruction, respectively. It is worth noting that the test dataset contains brain MR images, and HGGDP model is still trained on *SIAT* dataset.

First, the comparisons between MoDL and HGGDPRec are recoded in Table V and Fig. 12. The MoDL model is trained on 360 brain MR images of dataset shared by Ag-garwal *et.al.* The MRI data used for this study were acquired using a 3D T2 CUBE sequence with Cartesian readouts using a 12-channel head coil. The matrix dimensions were $256\times232\times208$ with 1 mm isotropic resolution. The coil sensitivity maps were estimated from the central k-space regions of each slice using ESPIRiT [57] and were assumed to be known during experiments. Thus, the data has dimensions in rows × columns × coils as $256\times232\times12$. The quantitative PSNR, SSIM and HFEN values for reconstruction results of the brain MR image Test1 under pseudo radial sampling at $R = 6$ is tabulated in Table V. It can be observed that the performance of MoDL is much inferior to the proposed HGGDPRec. Visual comparison of the reconstruction is shown in Fig. 12. It is not easy to find the difference between the reconstruction of MoDL and HGGDPRec with naked eye, but in the error map, our effect is better than MoDL.

Second, the comparisons between DDP and HGGDPRec are listed in Table VI and Fig. 13. Here, DDP [28] is trained on 790 (single coil, no phase) central T1 weighted slices with 1mm in-slice resolution from the HCP dataset and tested on an image from a volunteer acquired for this study (15 coils, complex image) along with the corresponding ESPIRiT coil maps. PSNR, SSIM and HFEN values of the final reconstruction results are listed in Table VI. It can be clearly seen that the reconstruction quality of HGGDPRec is superior to DDP. Particularly, as the undersampling rate increases, the performance gain becomes more striking.

To further prove the superiority of the HGGDP model, visualization results reconstructed by different methods on Test2 under 1D Cartesian sampling with two acceleration factors are provided in Fig. 13. It can be observed that the proposed model gets almost perfect reconstruction. HGGDPRec is able to fill in details to match the original image. In this case, the quality of the reconstructed image using DDP is better than the Zero-filled result, but due to the excessive smoothness of the image and the blur of the edge contour structure, some details are still lost.

TABLE V
RECONSTRUCTION PSNR, SSIM AND HFEN VALUES OF TEST IMAGE UNDER PSEUDO RADIAL SAMPLING AT $R = 6$ IN 12 COILS PARALLEL IMAGING.

| Method | $R = 6$ | | |
|---|---|---|---|
| | PSNR | SSIM | HFEN |
| Zero-filled | 25.94 | 0.7819 | 2.34 |
| MoDL | 39.75 | 0.9467 | 0.30 |
| HGGDPRec | **42.10** | **0.9717** | **0.17** |

TABLE VI
RECONSTRUCTION PSNR, SSIM AND HFEN VALUES OF TEST IMAGE AT 1D CARTESIAN UNDERSAMPLING PATTERNS IN 15 COILS PARALLEL IMAGING.

| Method | $R = 2$ | | | $R = 3$ | | |
|---|---|---|---|---|---|---|
| | PSNR | SSIM | HFEN | PSNR | SSIM | HFEN |
| Zero-filled | 30.55 | 0.8341 | 1.10 | 26.69 | 0.7455 | 1.94 |
| DDP | 37.31 | 0.9461 | 0.33 | 33.47 | 0.9063 | 0.80 |
| HGGDPRec | **39.69** | **0.9534** | **0.22** | **36.80** | **0.9374** | **0.34** |

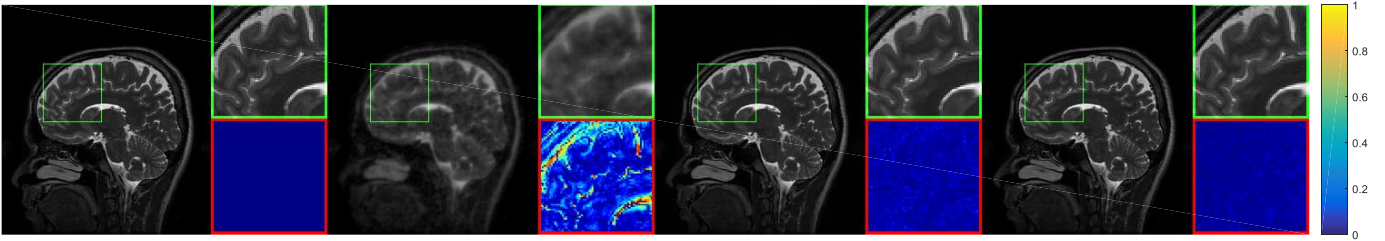

**Fig. 12.** Complex-valued reconstruction results on brain image at $R=6$ pseudo radial sampling in 12 coils parallel imaging. From left to right: Ground-truth, 6-fold pseudo radial undersample mask, reconstruction by Zero-filled, MoDL and HGGDPRec. Green and red boxes illustrate the zoom in results and error maps, respectively.

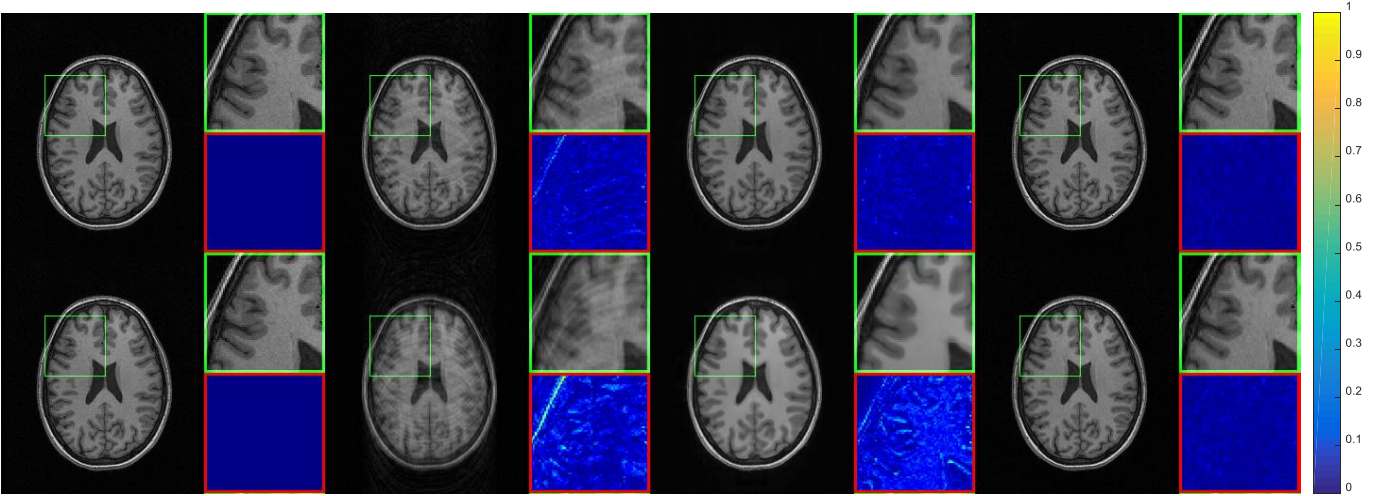

**Fig. 13.** Complex-valued reconstruction results on brain images at various 1D Cartesian undersampling percentages in 15 coils parallel imaging. From top to bottom: the reconstruction results on brain images at $R = 2$ and $R = 3$, respectively. From left to right: Ground-truth, various 1D Cartesian undersampling masks, reconstruction by Zero-filled, DDP and HGGDPRec. Green and red boxes illustrate the zoom in results and error maps, respectively.

### D. Ablation Study

We discuss the impact of several factors on HGGDPRec reconstruction: Size of $I$ in $\{\sigma_i\}_{i=1}^{I}$, number of channels, size of training patch, and initial values.

First, we use disturbance data with level random Gaussian noise as input in the HGGDPRec to train the network. Here, we set 4 groups of noises with $I = 8, 10, 12, 15$ as comparative experiments to study (eg the same proportional sequence with 1-0.01 but the different ratio). From Table VII, it can be seen that the PSNR, SSIM, and HFEN under $I = 10, 12, 15$ are obviously outperforms than $I = 8$. Especially, the result obtained by HGGDPRec becomes stable after 10. Perhaps with the increase of $I$, the result will be slightly improved, meanwhile, the calculation time cost will also increase linearly. Therefore, we set the number of $I$ to be 10.

Second, as stated in the Introduction section, multi-channel is a key contribution of our proposed method. Table VIII lists the performance of HGGDP under different channels at multi-channel training data. The effect presents a convex shape, which is considerable when the number of channels is 3. Simultaneously, as R increases, the difference between 3 channels and 2 channels becomes smaller, and the difference from 4 channels becomes larger. This phenomenon indicates that as the problem becomes more ill−posed, the strength of enforcing a multi-channel prior decrease.

Third, we investigate the influence of using different size patches of training data on HGGDP. Table IX lists the reconstruction results with patch sizes of 32,64,96. It is obvious that the effect of patch size of 32 is worse than that of 64,96. The phenomenon shows that too small patch size will destroy the structure of the image and make the reconstruction result poor. After it is larger than 64, the reconstruction result is relatively stable, and the difference between the block size of 64 and 96 is negligible. Meanwhile, considering that the larger the image block size, the greater the cost of occupying the GPU, the patch size of training data is set to be 64.

TABLE VII
PERFORMANCE OF RECONSTRUCTING 31 TEST IMAGES BY DIFFERENT SIZE OF $I$ AT RADIAL SAMPLING TRAJECTORIES IN SINGLE COIL.

| | $I$ | 8 | 10 | 12 | 15 |
|---|---|---|---|---|---|
| $R=4$ | PSNR | 34.87 | **35.55** | 35.47 | 35.48 |
| | SSIM | 0.9240 | **0.9287** | 0.9243 | 0.9233 |
| | HFEN | 0.59 | **0.50** | 0.51 | 0.50 |
| $R=10$ | PSNR | 29.70 | 30.85 | 30.77 | **30.92** |
| | SSIM | 0.8298 | **0.8488** | 0.8379 | 0.8385 |
| | HFEN | 1.56 | 1.29 | 1.32 | **1.28** |

TABLE VIII
PERFORMANCE OF RECONSTRUCTING 31 TEST IMAGES BY DIFFERENT NUMBER OF CHANNELS AT RADIAL SAMPLING TRAJECTORIES IN SINGLE COIL.

| | Channel number | 2 | 3 | 4 |
|---|---|---|---|---|
| $R=4$ | PSNR | 35.22 | **35.55** | 35.14 |
| | SSIM | 0.9277 | **0.9288** | 0.9260 |
| | HFEN | 0.51 | **0.5** | 0.52 |
| $R=10$ | PSNR | 30.69 | **30.88** | 30.37 |
| | SSIM | 0.8466 | **0.8488** | 0.8403 |
| | HFEN | 1.31 | **1.29** | 1.37 |

TABLE IX
PERFORMANCE OF RECONSTRUCTING 31 TEST IMAGES BY DIFFERENT SIZE OF TRAINING PATCHES AT RADIAL SAMPLING TRAJECTORIES IN SINGLE COIL.

| | Patch size | 32 | 64 | 96 |
|---|---|---|---|---|
| $R=4$ | PSNR | 34.93 | **35.55** | 35.48 |
| | SSIM | 0.9251 | **0.9288** | 0.9282 |
| | HFEN | 0.57 | **0.5** | 0.52 |
| $R=10$ | PSNR | 29.91 | 30.88 | **30.89** |
| | SSIM | 0.8345 | **0.8488** | 0.8485 |
| | HFEN | 1.49 | **1.29** | 1.3 |

Finally, we investigate how different initial values would affect the efficacy of the proposed method to reconstruct MR images. To this end, two different initializations are used respectively, namely initializing with the uniform noise $x \sim N(-1,1)$ and zero-filled data. Table X presents the results produced by our method with two initializations. The results gained by HGGDPRec are almost the same regardless of initializations, which empirically indicates that our method is insensitive to initialization.

As the cost time is an important factor in the reconstruction process, the iteration number $T$ of the inner loop is investigated. In Table XI, it can be found that the reconstruction quality has no obvious change after $T = 60$. As the acceleration factor $R$ increases, reconstruction with larger $T$ is needed to attain the best result.

TABLE X
PERFORMANCE OF RECONSTRUCTING 31 TEST IMAGES BY DIFFERENT INITIAL VALUES AT RADIAL SAMPLING TRAJECTORIES IN SINGLE COIL.

| Initial value | | $R=3.3$ | $R=4$ | $R=5$ | $R=10$ |
|---|---|---|---|---|---|
| Uniform noise $x \sim N(-1,1)$ | PSNR | **36.18** | **35.55** | **34.40** | **30.88** |
| | SSIM | **0.9364** | **0.9288** | **0.9130** | **0.8488** |
| | HFEN | **0.42** | **0.50** | **0.65** | **1.29** |
| Zero-filled | PSNR | 36.09 | 35.44 | 34.29 | 30.68 |
| | SSIM | 0.9307 | 0.9215 | 0.9032 | 0.8319 |
| | HFEN | 0.44 | 0.52 | 0.68 | 1.35 |

TABLE XI
PERFORMANCE OF RECONSTRUCTING 31 TEST IMAGES BY DIFFERENT SIZES OF $T$ AT RADIAL SAMPLING TRAJECTORIES IN SINGLE COIL.

| | | 20 | 40 | 60 | 80 | 100 |
|---|---|---|---|---|---|---|
| $R=4$ | PSNR | 34.79 | 35.37 | 35.51 | 35.54 | 35.55 |
| | SSIM | 0.9248 | 0.9276 | 0.9283 | 0.9286 | 0.9287 |
| | HFEN | 0.61 | 0.53 | 0.50 | 0.50 | 0.50 |
| $R=10$ | PSNR | 28.49 | 30.35 | 30.70 | 30.83 | 30.85 |
| | SSIM | 0.8011 | 0.8416 | 0.8459 | 0.8478 | 0.8488 |
| | HFEN | 1.94 | 1.43 | 1.34 | 1.30 | 1.29 |

## V. CONCLUSIONS

In this paper, a homotopic gradient of generative density prior was proposed. The generative modeling scheme first estimated the DSM and then employed Langevin diffusion for sampling. It leveraged the gradient of data density as prior and significantly improved the native NCSN for high diagnostic-quality image reconstruction. Two major characteristics were involved: Higher-dimensional and homotopic iteration. These factors jointly averted the data to be confined to low data density regions and avoided difficulties from the manifold hypothesis. Comprehensive experiment results demonstrated that HGGDPRec achieved superior performance. In the forthcoming future, more ways to form the high-dimensional space will be exploited and more imaging modalities will be applied to validate its efficiency, such as CT and PET reconstruction.